\documentstyle[preprint,aps]{revtex}
\begin{document}
\draft
%\preprint{\vbox{
%\hbox{IFT-P.059/93}
%\hbox{IFUSP/P-1075}
%\hbox{hep-ph/9310230}
%\hbox{August 1993}
%\hbox{revised Febreury 1994}
%}}
\title{
Radiatively induced electron and electron-neutrino masses}
\author{ F. Pisano$^a$, V. Pleitez$^a$ and M.D. Tonasse$^b$}
\address{
$^a$ Instituto de F\'\i sica Te\'orica\\
Universidade Estadual Paulista\\
Rua Pamplona, 145\\
01405-900-- S\~ao Paulo, SP\\
$^b$ Instituto de F\'\i sica da Universidade de S\~ao Paulo\\
01498-970 C.P. 20516-S\~ao Paulo, SP\\
Brazil}
\maketitle
\begin{abstract}
We consider, in the context of a 331 model with a single neutral
right-handed singlet, the generation of lepton
masses. At zeroth order two neutrinos and one charged lepton are
massless, while the other leptons, two neutrinos and two charged
leptons, are massive. However the charged ones are still mass
degenerate. The massless fields get a mass through radiative
corrections which also break the degeneracy in the
charged leptons.
\end{abstract}
\narrowtext
It is well known that in renormalizable theories, some masses or mass
differences vanish at tree level if there are some symmetries in
the theory which forbid them. This is maintained in higher order in
perturbation theory.
Or, if the respective higher order corrections are infinite the
introduction of the counter term necessary to remove the infinity
leaves these masses or mass differences as free parameters.
Notwithstanding, in theories with spontaneously broken symmetry if a
mass and mass counter term are forbidden by gauge structure, then
higher order corrections are finite and calculable~\cite{wggm}.
In this spirit, many mechanisms for finding fermion masses as
radiative corrections have been considered in the
literature~\cite{rev}. Here we will show how a mechanism
of this kind can be
implemented in the context of the recently proposed electroweak model
based on the $SU(3)_L\otimes U(1)_N$ gauge symmetry~\cite{pp,pf}.

In this model, leptons
are treated democratically with the three generations transforming as
$({\bf3},0)$ but with one quark generation (it does not matter which
one) transforming differently from the other two. This
condition arises because the model in order to be anomaly free must
contain the same number of triplets as antitriplets. Hence, the
number of generations is related to the number of quark colors.

In the minimal model the neutrinos remain massless since there is
a global symmetry which prevents them to get a mass. This
symmetry implies the conservation of the quantum number
${\cal F}=L+B$, where $L$ is the total lepton number $L=L_e+L_\mu+L_\tau$
and $B$ is the baryon number~\cite{pt}.
Here we will see that if we allow this symmetry to be explicitly
broken and also adding a single right-handed neutrino singlet,
one of the charged leptons and two neutrinos get mass
through radiative corrections.

Let us introduce  the
following Higgs scalars, $\eta=(\eta^0,\eta^-_1,\eta^+_2)^T$,
$\rho=(\rho^+,\rho^0,\rho^{++})^T$ and
$\chi=(\chi^- ,\chi^{--}, \chi^0)^T$
which transform, under $SU(3)_L\otimes U(1)_N$ as $({\bf3},0),({\bf3},1)$
and $({\bf3},-1)$, respectively.

The leptonic triplets are
$\psi_{aL}=(\nu''_a, l''_a, l''^c_a)^T\sim({\bf3},0)$,
where the double primed fields denote weak eigenstates,
$l''_a=e'',\mu'',\tau''$ and $\nu''_a=\nu''_e,\nu''_\mu,\nu''_\tau$.

The lepton mass term transforms as
${\bf3}\otimes{\bf3}={\bf3}^*_A\oplus{\bf6}_S$. Thus, we can introduce
a triplet, like $\eta$, or a symmetric antisextet $S=({\bf6}^*_S,0)$.
In the former case one of the charged leptons
remains massless, and the other two are mass degenerate. For this
reason it was chosen in Refs.~\cite{pf,fhpp} the latter one
in order to obtain arbitrary mass for charged leptons.

Here we will not introduce the sextet $S$ but only the triplets
$\eta,\rho$ and $\chi$; the respective VEV will be denoted by
$v_\eta,v_\rho$ and $v_\chi$.

The more general $SU(3)\otimes U(1)$ gauge invariant renormalizable
Higgs potential for the three triplets is
\begin{eqnarray}
V(\eta,\rho,\chi)&=&\mu_1^2\eta^\dagger\eta+\mu^2_2\rho^\dagger\rho+
\mu^2_3\chi^\dagger\chi+
\lambda_1(\eta^\dagger\eta)^2 \nonumber
\\ & &\mbox{}+\lambda_2(\rho^\dagger\rho)^2
+\lambda_3(\chi^\dagger\chi)^2
+(\eta^\dagger\eta)\left[\lambda_4(\rho^\dagger\rho )\right.
\nonumber \\ & &\mbox{}\left.+\lambda_5(\chi^\dagger\chi)\right]
+\lambda_6(\rho^\dagger\rho)(\chi^\dagger\chi)+
\lambda_7(\rho^\dagger\eta)(\eta^\dagger\rho)\nonumber
\\ & &\mbox{}  +
\lambda_8(\chi^\dagger\eta)(\eta^\dagger\chi)
+\lambda_9(\rho^\dagger\chi)(\chi^\dagger\rho)
\nonumber\\ & &\mbox{}+[\lambda_{10}(\eta^\dagger\chi)(\eta^\dagger\rho)
+\lambda'\epsilon^{ijk}\eta_i\rho_j\chi_k + H.c.].
\label{potential}
\end{eqnarray}
As we said before let us define the {\em
lepto-baryon} number ${\cal F}=L+B$, which is additively conserved.
As usually $B(l,\nu_l)=0$ for any lepton $l,\nu_l$, and $L(q)=0$ for
any quark $q$, ${\cal F}(l)={\cal F}(\nu_l)=+1$.
In order to make ${\cal F}$ a conserved quantum number in the Yukawa sector,
we also
assign to the scalar fields the following values $
-{\cal F}(\chi^-)=-{\cal F}(\eta^+_2)={\cal F}(\rho^{++})=
-{\cal F}(\chi^{--})=+2$,
and with all the other scalar fields carrying ${\cal F}=0$.

Notice that the ${\cal F}$-conservation forbids the quartic term
$\lambda_{10}(\eta^\dagger\chi)(\eta^\dagger\rho)$ in
Eq.~(\ref{potential})~\cite{laplata}. Hence, assuming that the
$\lambda_{10}$ term does exist we are violating explicitly the ${\cal F}$
symmetry. We must stress that if we had introduced the scalar sextet
and allow ${\cal F}$ to be broken, there will be additional terms involving
the sextet and the triplets as well. In this case it is not possible
to maintain neutrinos with calculable masses unless a fine tune is
imposed~\cite{pt}.

The $\lambda_{10}$ term has interactions
like $\rho^0\chi^0\eta^-_1\eta^+_2,\quad \eta^0\rho^0\eta^-_1\chi^+$,
and we have mixing between $\eta^-_1$ and $\eta^+_2$, etc. In fact, the
mass matrix in the singly charged scalars sector (in the
$\eta^-_1,\rho^-,\eta^-_2,\chi^-$ basis) is
\begin{equation}
v^2_\chi\left(
\begin{array}{cccc}
eba^{-1}-\lambda_7b^2\, &\, e-\lambda_7ab\, &\, \lambda_{10}b\, &\,
\lambda_{10}ab \\
e-\lambda_7ab\, &\, eab^{-1}-\lambda_7a^2\, &\, \lambda_{10}a\, &\,
\lambda_{10}a^2 \\
\lambda_{10}b \,&\, \lambda_{10}a\, &\,  eba^{-1}-\lambda_8\, &\,
eb-\lambda_8a \\
\lambda_{10}ab \,&\, \lambda_{10}a^2\, &\, eb-\lambda_8a & eab-\lambda_8a^2
\end{array}
\right),
\label{oba2}
\end{equation}
where $e=\lambda'/v_\chi$, $a=v_\eta/v_\chi$, $b=v_\rho/v_\chi$.
Hence we have a mixing among all singly charged scalars. The spectrum
from Eq.~(\ref{oba2}) will be given elsewhere but here we must stress that it
has two Goldstone
bosons. Notice that if there is no $\lambda_{10}$ term the mixing
occurs between $\eta_1^-,\rho^-$ and between $\eta^-_2,\chi^-$.

Since right-handed neutrinos transforming as singlets under
the gauge group do not contribute to the anomaly, their number is  not
constrained by the requirement of obtaining an anomaly free theory. Hence,
we can introduce, as in the standard electroweak model, an arbitrary
number of such fields. An interesting possibility is to introduce
just a single neutral singlet~\cite{cj}.

The Yukawa interaction in the leptonic sector plus a Majorana mass
term for the right-handed neutrino, is
\begin{equation}
{\cal
L}_{l\eta}=-\frac{i}{8}\sum_{a,b=e,\mu,\tau}
\epsilon^{ijk}F_{ab}\overline{(\psi_{Lai})^c}
\psi_{Lbj}\eta_k+\sum_a\hat{h}_a\overline{\psi_{aL}}\nu''_R\eta-\frac{1}{2}M
\overline{(\nu''_R)^c}\nu''_R
+H.c.
\label{yukawa}
\end{equation}
All the arbitrary constants in Eq.~(\ref{yukawa}) may be taken real and
positive.
The Yukawa couplings $F_{ab}$ must be antisymmetric due to Fermi
statistics. Explicitly we have
\begin{equation}
{\cal L}_{l\eta}=-i(v_\eta+\eta^0)\overline{l''_{Ra}}F_{ab}l''_{Lb}+
\frac{i}{2}\overline{l''_{Ra}}F_{ab}\nu''_{Lb}\eta^-_1
+\frac{i}{2}\overline{(\nu''^c)_{Ra}}F_{ab}l''_{Lb}\eta^+_2+H.c.,
\label{yukawa2}
\end{equation}
where
\begin{equation}
F_{ab}=\left(
\begin{array}{ccc}
0        \,&\,  -f_{e\mu} \,  &\, -f_{e\tau} \\
f_{e\mu}\, & \,      0 \,     &\, -f_{\mu\tau} \\
f_{e\tau}\, & \,  f_{\mu\tau}\, & \,    0     \\
\end{array}
\right).
\label{m1}
\end{equation}
The mass spectrum of the charge leptons is $0,m,-m$. We  can always define
$e''$ as the state with zero mass.
That is, we can choose a basis in which $f_{e\mu}=f_{e\tau}=0$. In
this case we have $m=v_\eta
f^2_{\mu\tau}$. In this basis the matrix in (\ref{m1}) can be
diagonalized  by an unitary matrix
\begin{equation}
U=\left(
\begin{array}{ccc}
1  & 0 & 0 \\
0  & \frac{1}{\sqrt2} & \frac{1}{\sqrt2} \\
0 & -\frac{i}{\sqrt2} & \frac{i}{\sqrt2}\\
\end{array}
\right).
\label{u}
\end{equation}
Notice that one of the mass is negative.  To get positive mass
eigenvalues we let $m$ be positive and redefine the respective field
with a $\gamma^5$ factor, i.e., $\tau'\to\gamma^5\tau'$.
Because of this $\gamma^5$ the $CP$ of $\tau'$ is $-1$.

The neutrinos mass term is
\begin{equation}
{\cal
L}_\nu=-\sum_{a=e,\mu,\tau}
h_a\bar\nu'_{La}\nu'_R-\frac{1}{2}M\overline{(\nu'_R)^c} \nu'_R+H.c.,
\label{cecilia}
\end{equation}
where $h_a=v_\eta\hat h_a$, and $\hat h_a$ are arbitrary dimensionless
parameters.  We can write the
mass term as $-\frac{1}{2}\bar N''M^\nu N''^c$ with $N''=(\nu''_{eL},
\nu''_{\mu L},\nu''_{\tau  L}, {\nu''_R}^c)^T$ and
\begin{equation}
M^\nu=\left(
\begin{array}{cccc}
0\,   & \, 0 \,     & \, 0  \,      &\, h_e \\
0\,   & \, 0 \, & \, 0 \,       & \, h_\mu \\
0 \,  &\,  0 \,     & \, 0 \,   & \, h_\tau \\
h_e\, &\, h_\mu \,   & \, h_\tau \,      & \,M
\end{array}
\right)
\label{mnu}
\end{equation}
We  can diagonalize the neutrino mass matrix by making $N'=\Phi PN''$ with
$P$ an orthogonal matrix,
\begin{equation}
P=\left(
\begin{array}{cccc}
h_\mu/(A^2-h_\tau^2)^{\frac{1}{2}} & h_e/(A^2-h_\tau^2)^{\frac{1}{2}}
               &    0   &  0   \\
h_eh_\mu h_\tau/(A^2-h_e^2)^{\frac{1}{2}}&
h_\tau h_\mu^2/(A^2-h_\tau^2)(A^2-h_e^2)^{\frac{1}{2}}
& -h_\mu/(A^2-h_e^2)^{\frac{1}{2}}& 0   \\
(h_em'_{\nu_P}-A^2)/D_1 &
h_\mu M/D_1  &
h_\tau M/D_1 & h_e M/D_1 \\
h_e(h_em'_{\nu_F}-A^2)/D_2 &
h_\mu(h_em'_{\nu_F}-A^2)/D_2 &
h_\tau(h_em'_{\nu_F}-A^2)/D_2 & -AM(h_em'_{\nu_F}-A^2)/D_2 \\
\end{array}
\right)
\label{nu1}
\end{equation}
where
$$
D_1^2=(A^2-h_\tau^2)^2+M^2A^2+(A^2-h_\tau^2-h_em'_{\nu_P})
(h_\tau^2-h_em'_{\nu_P})
$$
$$
D_2^2=A(A^2-h_\tau^2)(A^2-h_\tau^2+M^2-2h_em'_{\nu_F})
+h_e^2({m'}_{\nu_F}^2+2h_\tau^2).
$$
$N'=(\nu'_{1L},
\nu'_{2L},\nu'_{PL},\nu'_{FL})^T$, $A^2=h_e^2+h_\mu^2+h_\tau^2$ and
$\Phi$ is a diagonal
phase matrix $\Phi=diag(1,1,i,1)$. At this stage, the neutrino
mass spectrum consists of two massless fields $\nu'_{1,2}$ and two
Majorana massive $\nu'_{P,F}$ neutrinos~\cite{cj}
\begin{equation}
m'_{\nu_P}=\frac{1}{2}[(4A^2+M^2)^{\frac{1}{2}}-M],\;\;
 m'_{\nu_F}=\frac{1}{2}[(4A^2+M^2)^{\frac{1}{2}}+M],
\label{fp}
\end{equation}

Note that $m'_{\nu_F}$ is arbitrary and in fact could be heavier than the
lepton-$\tau$. Constraints on the masses $m'_{\nu_P}$ and $m'_{\nu_F}$
coming from the measured $Z^0$ invisible width were considered in
Ref.~\cite{eovr}.

In  the primed basis for the charged leptons, Yukawa couplings with
the scalars $\eta^-_1$ and $\eta^+_2$ can be written as
\begin{equation}
\frac{i}{2}\overline{l'_{Ra}}\left(U F\right)_{ab}
\nu''_{Lb}\eta^-_1
+\frac{i}{2}\overline{(\nu''^c)_{Ra}}(FU^\dagger)_{ab}l'_{Lb}\eta^+_2+H.c.
\label{yukawac}
\end{equation}
As there are four neutrinos but only three charged leptons, it is possible
to extend the charged lepton column with a zero on the fourth row in such
a way that in Eq.~(\ref{yukawac}) all matrices are $4\times 4$.
In Eq.~(\ref{yukawac}) the neutrinos are still linear combinations of
the mass eigenstates, $\nu''_{a L}=(-\Phi P^T)_{al}N'_{lL},\,l=1,2,3,4$.
Let us denote $\Gamma= UF $. Notice that in Eq.~(\ref{yukawac}) the
$F$ matrix is in the  basis in  which $f_{e\mu}=f_{e\tau}=0$ and for
this reason the electron does not interact with any neutrino.

Due to the mixing of  $\eta^-_1$ and $\eta^-_2$ and to the
non-diagonal Yukawa couplings $\Gamma$,
diagrams like the one
showed in Fig. 1 exist and they are finite.  Notice that the primed fields
$(\mu',\tau')$ couple with the two massive neutrinos $\nu'_P,\nu'_F$,
then the  neutrino
masses insertions are $m'_{\nu_P},m'_{\nu_F}$.
Then, the diagrams in Fig. 1 induce a contribution to the mass
matrix of the $\mu',\tau'$. In fact contributions from Fig. 1 have
the following form
\begin{equation}
\delta_{ab}\sim \lambda_{10}m'\alpha_l\beta_{l'}
\Gamma_{al}\Gamma_{l'b}\left(\frac{v_\rho v_\eta}{m_\eta^2}\right)
\ln\left(
\frac{m_{\eta_1}^2}{m_{\eta_2}^2}\right),\quad a,b=\mu,\tau
\label{lm}
\end{equation}
where $m'$ means $m'_{\nu_P}$ or $m'_{\nu_F}$,  $m_{\eta_1}^2,
m_{\eta_2}^2$ are typical masses in the scalar
sector and $m^2_\eta$ is the greatest of them. The $\Gamma$'s are
the couplings appearing in Eq.~(\ref{yukawac}) and $\alpha_l,\beta_{l'}$
denote the $\nu''$'s projections in the $\nu'_{P,F}$ components i.e.,
can be read off from $\nu''_{aL}=-(\Phi P^T)_{al}N_l,\;l=3,4$.
As an example the $\nu'_F$ contribution in Fig. 1 induces the  following
mass matrix in the $\mu',\tau'$ basis
\begin{equation}
m'\alpha_l\beta_{l'}
\Gamma_{al}\Gamma_{l'b}\sim\frac{1}{D_2^2}\left(
\begin{array}{cc}
mD_2^2+m'_{\nu_F}(h_em'_{\nu_F}-A^2)^2(h_\tau-h_\mu)^2 &
-im'_{\nu_F}(h_em'_{\nu_F}-A^2)^2(h_\tau^2-h_\mu^2) \\
im'_{\nu_F}(h_em'_{\nu_F}-A^2)^2(h_\tau^2-h_\mu^2)
&mD_2^2+m'_{\nu_F}(h_em'_{\nu_F}-A^2)^2(h_\tau+h_\mu)^2
\end{array}
\right).
\label{mutau}
\end{equation}
Notice that the diagonal terms have the contributions of the tree level
mass $m$. We see from Fig.1 that an arbitrary mass matrix arises for the
charged leptons $\mu'$ and $\tau'$ at the 1-loop level breaking their
mass degeneracy. Since $\nu'_F$ can be heavier than the tau lepton~\cite{eovr}
the mass difference $m_\tau-m_\mu$ may be fitted with
reasonable values for the parameters present in the model. Notwithstanding,
the electron is still massless.

Since at the tree level $\nu'_1$ and $\nu'_2$ are massless we may define linear
combinations, say $\tilde\nu'_1$ and $\tilde\nu'_2$ in such a way that
$\tilde\nu'_2$ does not couple to one of the charged leptons, say the
electron.
However, if one makes this choice in the vector current, $\tilde\nu'_2$ will
still couple to the electron through the Yukawa interactions in
Eq.~(\ref{yukawac}). This is due to the presence of the $F$ matrix. On the
contrary if we choose that $\tilde\nu'_2$ does not couple to the electron in
Eq.~(\ref{yukawac}) it will couple to it through the vector current.

At 1-loop level processes like those showed in Fig.2 are also
possible. Here the mass insertions are $m$ i.e., the charged lepton
mass at tree level. It implies also a symmetric matrix in addition to
$M^\nu$ in Eq.~(\ref{mnu}). Hence, the massless neutrinos (at tree
level)  $\nu'_2$ acquire masses from 1-loop radiative corrections.
In Fig. 2 we show the case of $\nu'_2\to\nu'_F$ mixing. However, $\nu'_1$
is still massless.

After these loop corrections have been taken into account we have
the basis $(e'',\mu,\tau)$ and $(\nu'_1,\nu_2,\nu_P,\nu_F)$ with the
respective masses $(0,m_\mu,m_\tau)$, and $(0,m_2,m_{\nu_P},m_{\nu_F})$.

In the model there are two singly charged vector bosons $W^\pm$
and $V^\pm$. Their interactions with the leptons are
$\bar\nu_{lL}V_l\gamma^\mu
l_LW^+_\mu$ and $\bar l^c_LV^\dagger_l\gamma^\mu
\nu_{lL}V^+_\mu$~\cite{pp} respectively.
At this stage since three neutrinos are already massive we have a
general mixing among them i.e., $ V_l= P'^T\Phi P^TU^\dagger U'^\dagger$
is a $4\times 4$ matrix, $P'$ is the arbitrary unitary $3\times 3$
matrix which diagonalize the mass matrix of the
three massive neutrinos after the contributions
of Fig. 2 have been taken into account. On the other hand $U'$ has
the same structure than $U$ in Eq.~(\ref{u}) but now
with arbitrary elements. The matrices $U,U'$ and $P'$
are appropriately extended by adding zeros and with $U_{44}=U'_{44}=P'_{11}=1$,
in order to write them as $4\times 4$ matrices and the charged lepton
column written as $(e'',\mu,\tau,0)$. Hence, it is straightforward to
convince ourselves that $V_l$ has the appropriate form to induce
diagrams like that showed in Fig. 3. These diagrams which exchange one
$W^-$ and one $V^+$ as in Ref.~\cite{babu} are possible and they are
responsible for the mass of the electron as is shown in Fig.3. In
this figure, we show only the contribution involving $\nu_F$
and the tau lepton. This is a higher order
contribution, however it is proportional
to $m_\tau m_{\nu_F}^2$ and for this reason
the electron mass can be of the appropriate size.

Let us back to the $\nu'_1$ neutrino. This particle is up to now massless,
however notice that after the electron has got a mass there is a contribution
similar to the Fig. 2 but now the electron in the internal line. This kind of
processes mix all the four neutrinos and $\nu'_1$ acquire a mass.

We have shown that if one does not introduce the sextet $S$, it is
possible to give the right mass to all leptons if at least a
single right-handed neutrino and a ${\cal F}$-violating term in the
scalar potential are added to the minimal model. It is possible, of
course, the case with three right-handed  neutrinos, but we will not
treat this case here.

It is interesting that
in a supersymmetric version of the model it is also possible to give mass
to the charged leptons without introducing the sextet of scalars~\cite{ema}.

Hence, in this 331 model, the smallness of the masses of
the electron, the lightest neutrinos and the  mass difference
between muon and tau arise naturally.

We are very grateful to Funda\c c\~ao de Amparo \`a Pesquisa do Estado de
S\~ao Paulo (FAPESP)  for full financial support (M.D.T.) and
Con\-se\-lho Na\-cio\-nal de De\-sen\-vol\-vi\-men\-to Cien\-t\'\i
\-fi\-co e Tec\-no\-l\'o\-gi\-co (CNPq) for partial (V.P.) and full (F.P.)
financial support. We also thank O.L.G. Peres for useful discussions.

%%%%%%%%%%%%%%%%%%%%%%%%%%%%%%%%%%%%%%%%%%%%%%%%%%%%%%%%%%%%
%%%%%%%%%%%%%%%%%%%%%%% FIGURES %%%%%%%%%%%%%%%%%%%%%%%%%%%%
%%%%%%%%%%%%%%%%%%%%%%%%%%%%%%%%%%%%%%%%%%%%%%%%%%%%%%%%%%%%
\begin{figure}[p]
%\postscript{ppt1.ps}
\caption
{One-loop contributions to the charged lepton ($\mu',\tau'$) mass matrix.
With this contribution the mass matrix appers in Eq.(13). There is a similar
diagram with $\nu'_P$ on the internal line.}
\label{fig1}
\end{figure}
\begin{figure}[p]
%\postscript{ppt2.ps}
\caption
{One-loop diagram inducing a mass for the $\nu'_2$ neutrino.}
\label{fig2}
\end{figure}
\begin{figure}[p]
%\postscript{ppt3.ps}
\caption
{ Diagram inducing a mass for the electron.}
\label{fig3}
\end{figure}

\end{document}